\documentstyle[11pt,newpasp,twoside,epsf]{article}
\def\edcomment#1{\iffalse\marginpar{\raggedright\sl#1\/}\else\relax\fi}
\reversemarginpar

\begin{document}
\title{Preliminary Observational Results of Tidal Synchronization
in Detached Solar-Type Binary Stars}
\author{S{\o}ren Meibom}
\affil{University of Wisconsin - Madison, Wisconsin, USA}

\begin{abstract}
We present preliminary observational results on tidal synchronization
in detached solar-type binary stars in the open clusters M35 (NGC2168;
$\sim$ 150 Myr) and M34 (NGC1039; $\sim$ 250 Myr). M35 and M34 provide
populations of close late-type binaries with ages that make them attractive
observational tests of models of tidal synchronization during the early
main-sequence phase. A combined dataset of stellar rotation periods
from time-series photometry and binary orbital periods and eccentricities
from time-series spectroscopy enables us to determine the angular rotation
velocity of the primary star and the orbital angular velocity at periastron.
Comparison of the stellar and orbital angular velocities provides
information about the level of synchronization in individual binary
stars.
\end{abstract}

\section{Introduction}

The evolution of stellar angular momentum remains a challenge
to our understanding of star formation and early stellar evolution.
Angular momentum in solar-type stars is not conserved at any time
from birth until death. Long-standing research has demonstrated
angular momentum loss via winds, and recently the roles of proto-stellar
disks and jets have come to the fore. However, few studies have
confronted the role of binary companions, despite the fact that
most stars form and evolve in binary systems (Larson 2002; Duquennoy
\& Mayor 1991).

Tidal forces in close detached binary stars drive the exchange
of angular momentum between the stars and their orbital motion
(Witte \& Savonije 2002; Zahn 1989,1977; Hut 1981). Depending
on the strength of the tidal interactions the stars in a binary
system will synchronize their rotation with the orbital motion.
The rate of synchronization depends sensitively on the separation
of the two stars and on the mechanism for dissipation of kinetic
energy into heat within the stars. Understanding the process
of tidal dissipation will allow important information about
a binary's past angular momentum evolution to be derived, and
possibly constrain the conditions of its formation.

The amount of observational data suitable for testing the rate
of tidal synchronization in late-type close binaries is sparse.
Most of the observational work on synchronization in close binaries
has concentrated on early-type stars with radiative envelopes
where the mechanism for tidal dissipation is different (Claret
\& Cunha 1997; Giuricin, Mardirossian, \& Mezzetti 1984).

Efforts to theoretically model the evolution of tidal synchronization
in late-type stars are ongoing. The main difference between current
models lies in the mechanism by which kinetic energy is being dissipated
within the stars (the tidal dissipation mechanism).
The equilibrium tide theory describes the retardation
of the hydrostatic tidal bulge (the equilibrium tide) due to the coupling
of the tidal flow to the motion of turbulent eddies in the stellar
convective envelope (Zahn 1989;1977; Hut 1981). The dynamical tide
theory describes the excitation, damping and resonances of gravity
(g) waves in the radiative zones of stars due to the tidal forcing
by the companion star (Savonije \& Papaloizou 1983; Zahn 1977).
Given an energy dissipation model, tidal theory makes explicit predictions
for the evolution of stellar angular momentum in a binary system.

In Figure 1 Zahn \& Bouchet (1989; hereinafter ZB) use the
equilibrium tide theory to predict the rotational evolution of
a solar-mass pair with an initial orbital period of 5 days and
an initial orbital eccentricity of 0.3 (dashed curve).
Synchronization is rapidly achieved near the stellar birth-line,
maintained until shortly after 1 Myr, then lost as the contractions
of the stars reduce the tidal forces and spin up the stars.
These super-synchronous stars do not regain synchronization
until $10^9$ yr. Similar modeling predicts pseudo-synchronization
with the periastron angular velocity in longer-period eccentric
orbits (cf. Hut 1981).

\begin{figure}[t!]  
\plotone{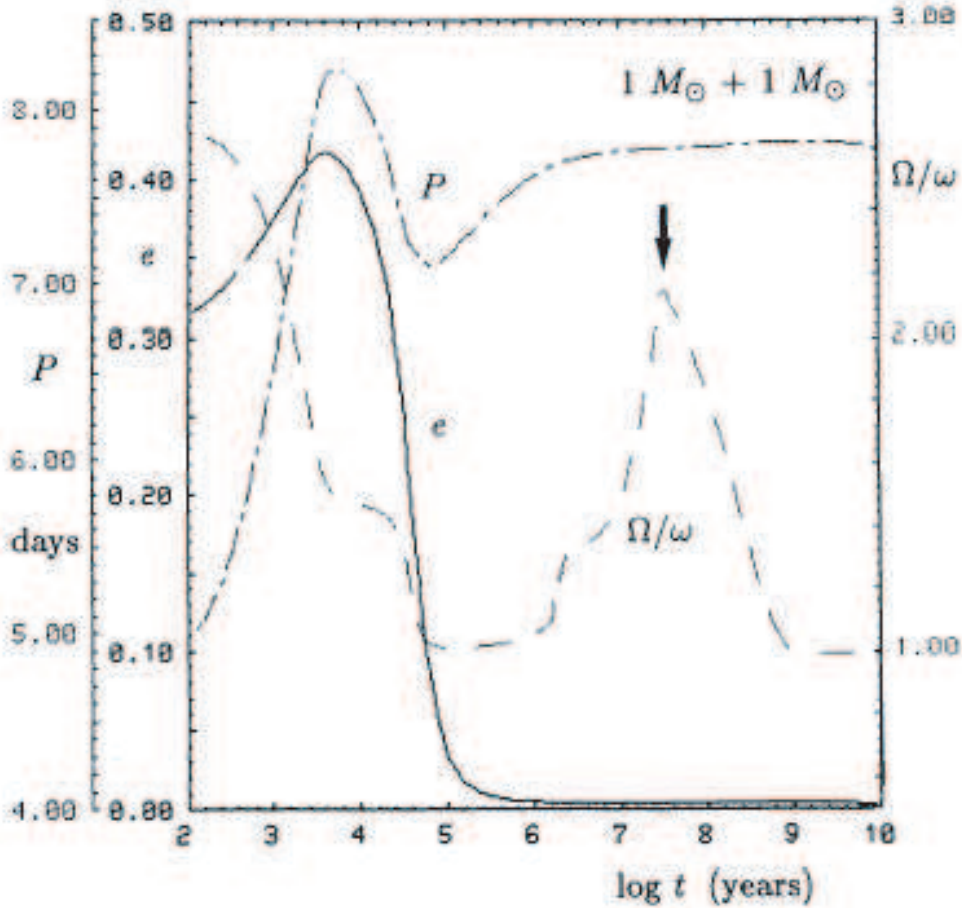} 
\caption{Evolution in time of the ratio between the rotational
and orbital velocities ($\Omega/\omega$; dashed curve) for a
binary with solar-mass components, an initial orbital period
of 5 days, and an initial orbital eccentricity of 0.3 (Zahn \& Bouchet 1989).}
\end{figure}

Figure 2 shows the prediction by Witte \& Savonije (2002;
hereinafter WS) of the rotational evolution of a solar-mass
pair with an initial orbital period of 16 days and an initial
orbital eccentricity of 0.6. WS applies the dynamical tide theory
with inclusion of resonance locking. The stars' initial rotational
speed are 20\% of their breakup speed, corresponding to a period
of 12 hours. Strong retrograde torques rapidly spin down the stars
during the first few tens of millions of years ($t \sim$ 0.08 Gyr),
after which there is a reduction in the stellar spin-down rate
lasting to $t \sim$ 0.5 Gyr. At that time stellar rotation is
approximately synchronous to the orbital angular velocity at
periastron. Spin-up to $\sim 1.4~\omega_{per}$ occurs from
$t \sim$ 0.5 - 0.9 Gyr due to prograde forcing torques, at
which point a decrease in stellar separation triggers efficient
retrograde forcing of the stars resulting in efficient spin-down
to sub-synchronous rotation at $t \sim$ 3 Gyr. As the binary
orbit circularizes the retrograde forcing diminishes and
the stars approach synchronous rotation over the remainder
of the main-sequence phase.

\begin{figure}[t!]
\plotone{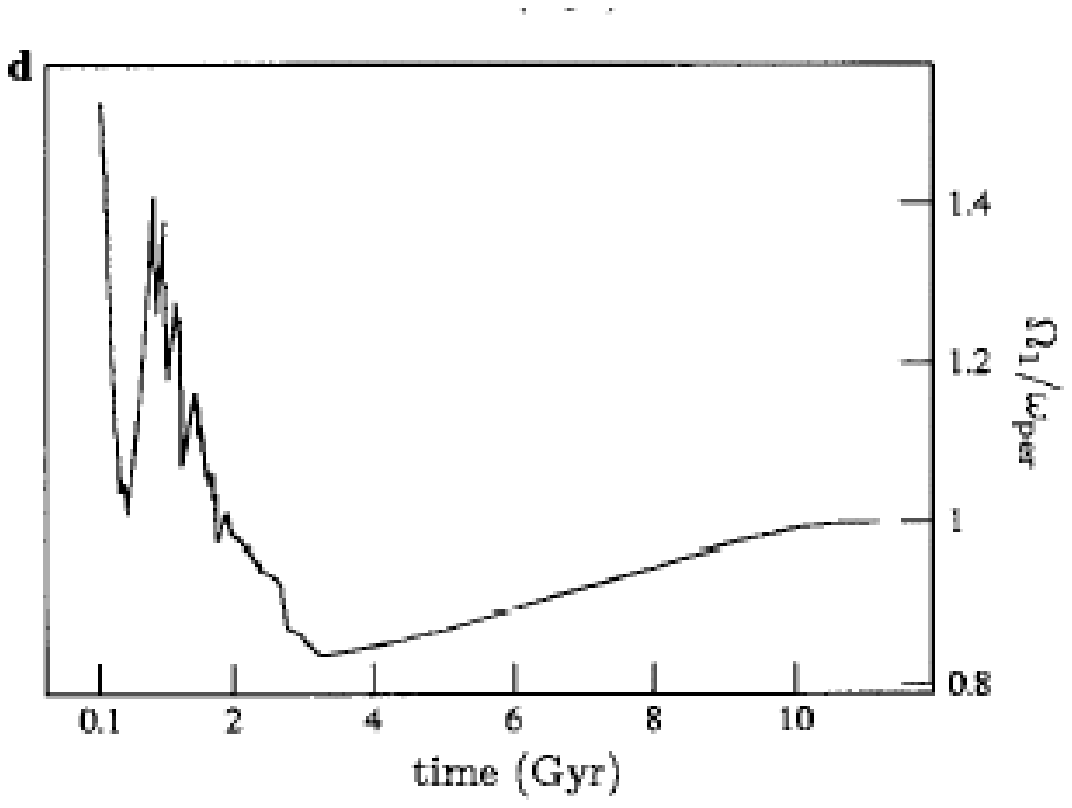} 
\caption{The predicted rotational evolution of the solar-mass
components in a close binary with an initial orbital period
of 16 days and an initial orbital eccentricity of 0.6 (Witte
\& Savonije 2002). The stellar angular rotational velocity
($\Omega_1$) is shown relative to the orbital angular velocity
at periastron ($\omega_{per}$). The modeling starts at the ZAMS.}
\end{figure}

Note that the model by ZB includes the pre main-sequence (PMS)
evolutionary phase, whereas the WS modeling starts at the ZAMS
(Zero Age Main-Sequence). The success of either of these theoretical
models can be measured by their ability to predict the observed
rotational evolution of stars in close binaries with known ages.

\section{Observational Program and Goal}

Claret \& Cunha (1997) analyzed the validity of Zahn's theories
of tidal synchronization using the data from Andersen (1991).
This paper is so far the only published study found by us to
include late-type stars. Therefore, there is a need for large
coeval samples of late-type binaries with accurate information
about the orbital angular velocity as well as the rotational
angular velocity of the primary star.

The combination of observed stellar rotation periods and orbital
periods and eccentricities for binaries of known ages will directly
test the theoretical predictions, and thereby the energy dissipation
rates in stellar interiors. Inspired by the success of the distribution
of orbital eccentricities versus orbital periods (the $e-\log(P)$
diagram) to constrain theories of tidal circularization (see Mathieu
this volume, Meibom \& Mathieu 2004, and references therein), we
intend to start populating a similar diagram, the $\Omega/\omega-\log(P)$
diagram, with the purpose of constraining theories of tidal synchronization.
The $\Omega/\omega-\log(P)$ diagram represents, as a function of binary
period, the ratio of stellar angular velocity of a binary's primary
star ($\Omega$) to the orbital angular velocity at periastron ($\omega$).

To obtain that observational goal we have conducted two parallel
observational programs on the open clusters M35 (NGC2168; $\sim$
150 Myr) and M34 (NGC1039; $\sim$ 250 Myr):
{\bf 1)} High precision radial-velocity surveys to identify
binaries and determine their orbital parameters. The surveys
are carried out using the WIYN 3.5m telescope, Kitt Peak, Arizona,
equipped with the Hydra Multi Object Spectrograph. The surveys
are part of the WIYN Open Cluster Study (WOCS) radial-velocity
survey program (Mathieu 2000). Radial velocities with measurement
accuracies of $< 0.5$ km/s are achieved to V = 17 in 2-hour
integrations for late-type stars with narrow lines. The selected
brightness range in the two clusters $(12 < V < 17)$ correspond
to stellar masses from $\sim 1.5$ to $\sim 0.5$ solar masses.
At the present time, orbital parameters have been determined
for 32 spectroscopic binaries in M35 and 18 in M34.
{\bf 2)} Comprehensive photometric time-series surveys to
determine stellar rotation periods from light modulation by
starspots on the surfaces of late-type stars.
The time-series surveys are based on synoptic data with
a frequency of one observation per night from October 2003
till March 2004, and on 2 weeks of 5-6 observations per night
in December 2003. The photometric data presented for M35
have been obtained using the WIYN 0.9m telescope at
Kitt Peak, Arizona. We have not yet determined rotation
periods for stars in M34 but present here results from a
similar study by Barnes (private comm.) using the
Hall 42-inch telescope at Lowell Observatory, Arizona.

\section{Preliminary Results}

To populate the $\Omega/\omega-\log(P)$ diagram we are dependent
on being able to determine the rotation period of the primary
star in binaries for which we also have information about the
orbital period and eccentricity. We show in Figure 3 through
6 the preliminary observational results for 4 such binary systems
in M35. Relevant orbital and rotational parameters are listed in
the figure captions.

Binary 6821 (Figure 3): This single-lined spectroscopic binary
(SB1) member of M35 has a circular orbit with a 2.25-day period. 
We estimate the mass of the primary component to be $0.8~M_{\odot}$
based on a fit of the Yale (Y2) stellar evolutionary models
(Yi et al. 2003) to the cluster sequence. The orbital parameters
and solution are shown on the left in Figure 3. From the photometric
data we found that the integrated stellar brightness of binary 6821
varied with a period of 2.3 days. The phased photometric light-curve
is shown on the right in Figure 3, together with the resulting
power-spectrum and the unphased light-curve. The similarity of the
orbital period and the period of variability in the photometric
data suggest that the rotation of the primary star in binary
6821 is synchronized to the orbital motion. The derived ratio
of stellar angular velocity to the orbital angular velocity 
($\Omega/\omega$) is 0.94.

\begin{figure}[t!]
\plottwo{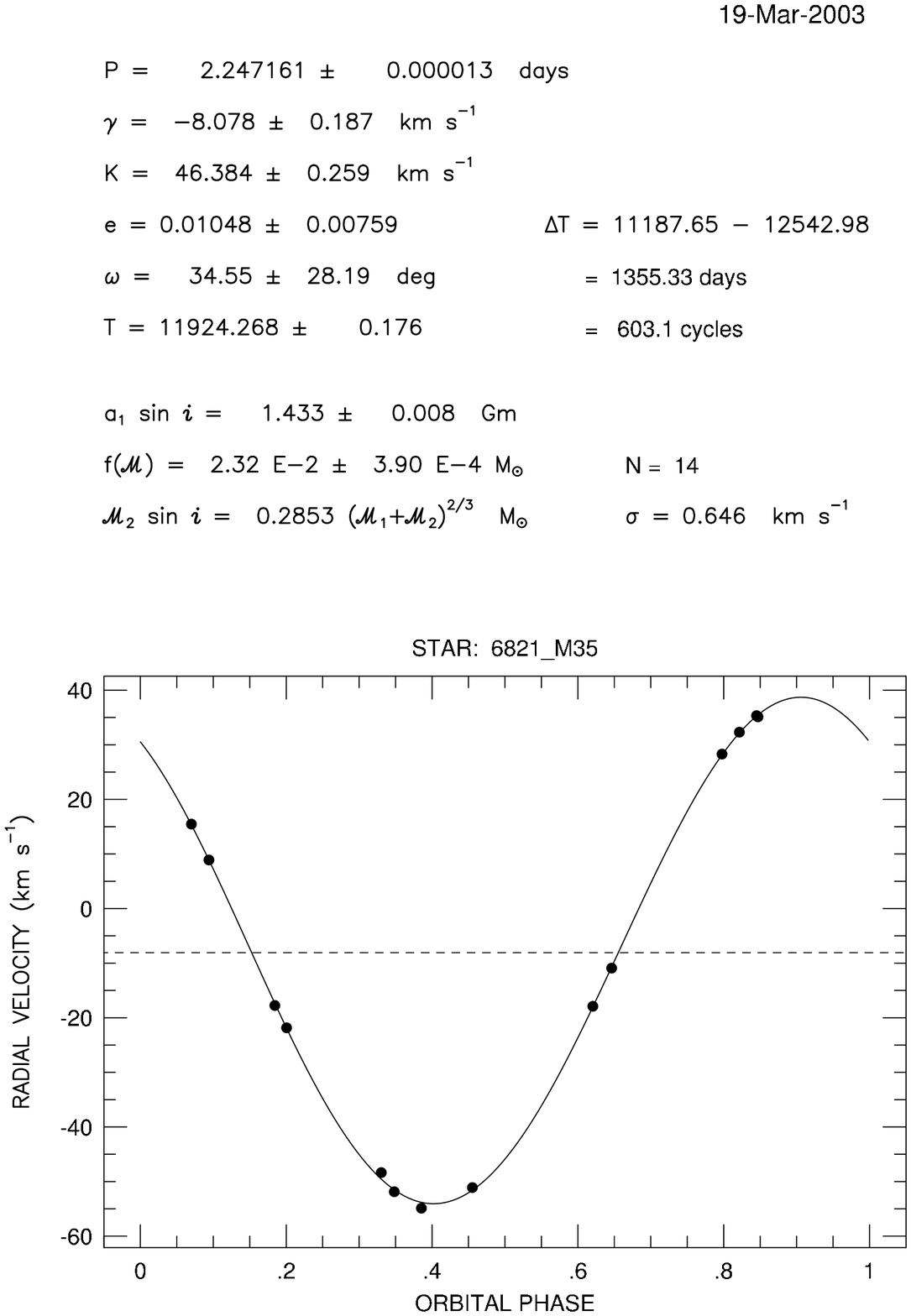}{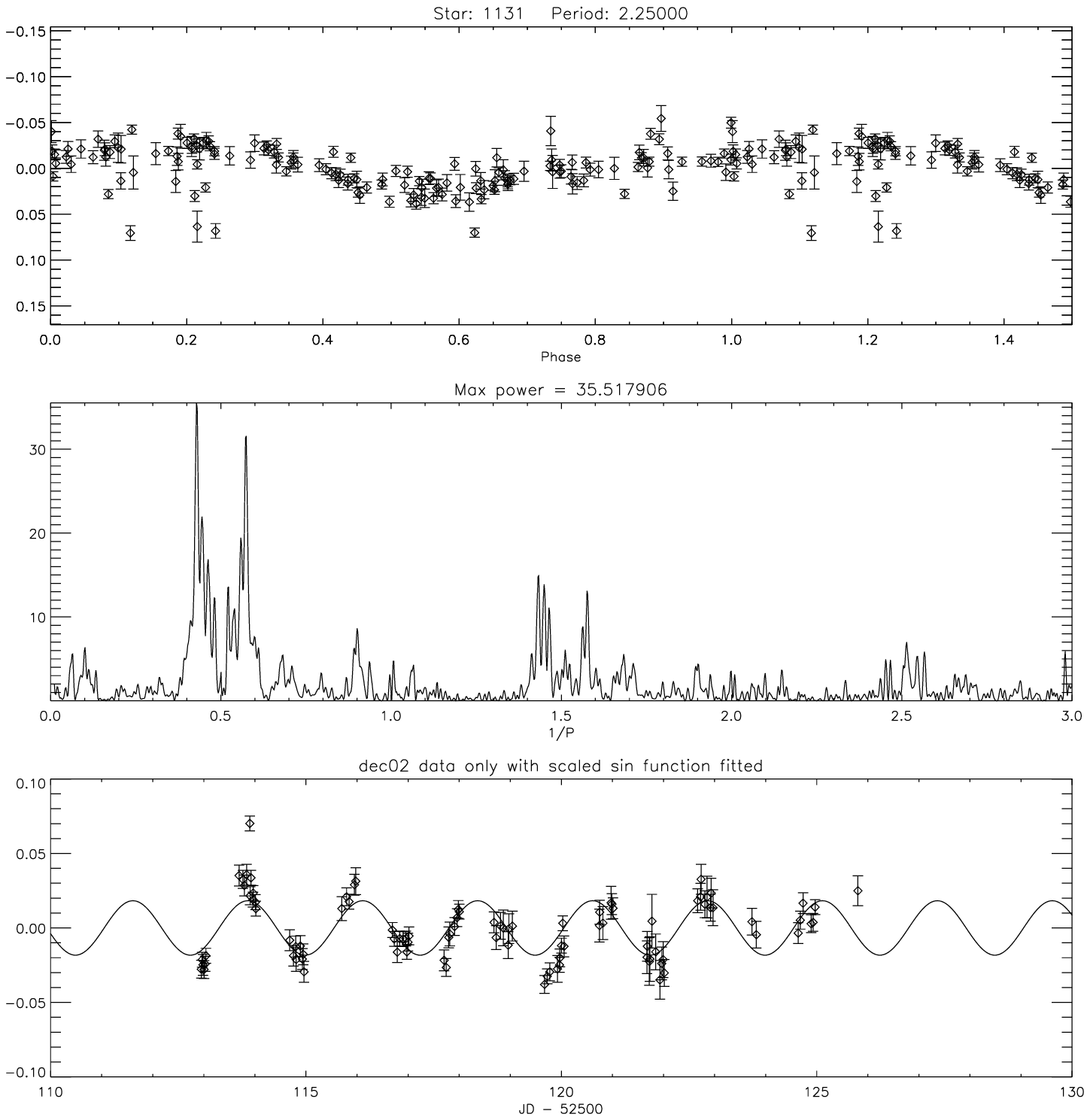}
\caption{{\bf Left:} The orbital parameters and phased radial-
velocity curve for binary 6821. The best fit orbit has an eccentricity
of 0.01 and a period of 2.25 days. {\bf Right:} The photometric
light-curve for binary 6821 phased to a period of 2.3 days together
with the power spectrum in frequency space and the unphased light-curve.
A sine curve with a period of 2.3 days has been overplotted on
the unphased photometric data.}
\end{figure}

Binary 774 (Figure 4): This SB1 has a circular orbit with a
period of 10.3 days. The estimated mass of the primary component
is $0.95~M_{\odot}$. The orbital parameters and solution are
shown on the left in Figure 4. The photometric data show 10.1-day
periodic variability in the integrated stellar brightness
of binary 774. The phased photometric light-curve is shown on
the right in Figure 4 together with the resulting power-spectrum
and the unphased light-curve. Again, the similarity of the orbital
period and the period of variability in the photometric
data suggest that the rotation of the primary star in binary
774 is synchronized to the orbital motion. The derived ratio
of stellar angular velocity to the orbital angular velocity
($\Omega/\omega$) is 0.99.
 
\begin{figure}[t!]
\plottwo{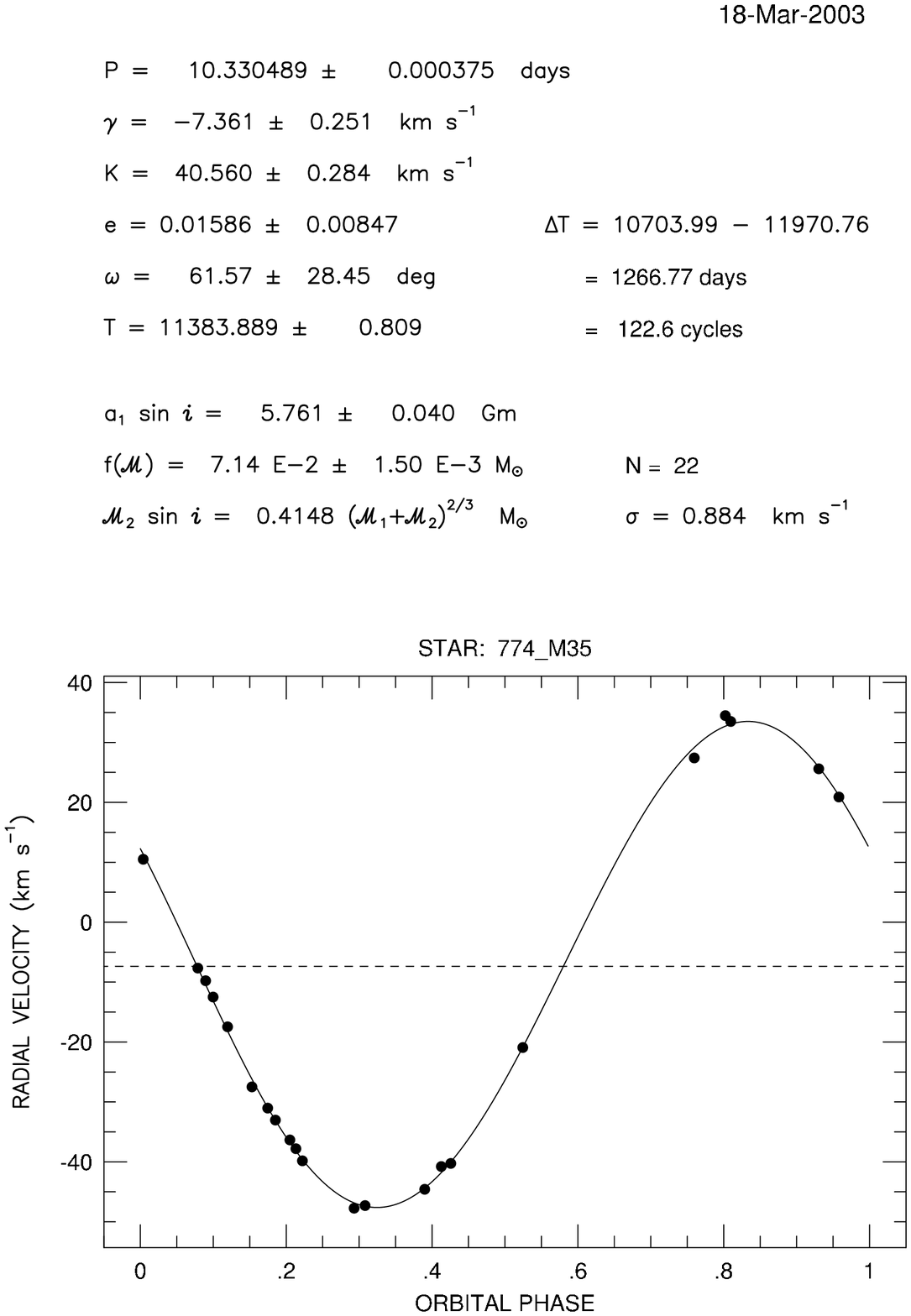}{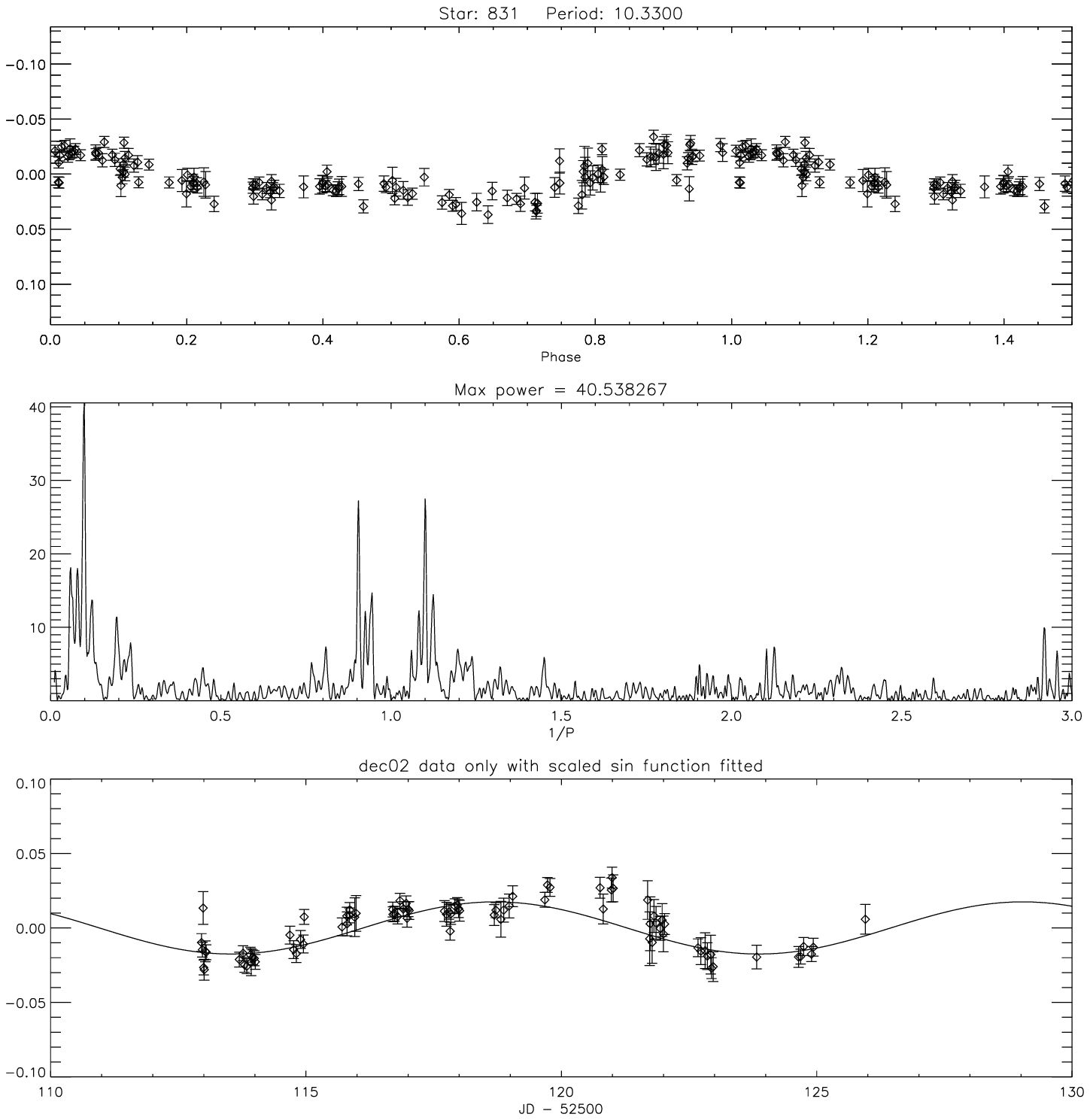}
\caption{{\bf Left:} The orbital parameters and phased radial 
velocity curve for binary 774. The best fit orbit has an eccentricity
of 0.01 and a period of 10.3 days. {\bf Right:} The photometric
light-curve for binary 774 phased to a period of 10.1 days together
with the power spectrum in frequency space and the unphased light-curve.
A sine curve with a period of 10.1 days has been overplotted
the unphased photometric data.}
\end{figure}

Binary 3121 (Figure 5): This SB1 has an eccentric orbit
($e = 0.27$) with a period of 30.1 days. The estimated mass
of the primary component is $0.95~M_{\odot}$. The photometric
data show a 2.8-day periodic variability in the integrated
stellar brightness of the binary. The phased photometric
light-curve is shown on the right in Figure 5. The period
of variability in the photometric data thus suggest that
the rotation of the primary star in binary 3121 is super-synchronous.
The ratio of stellar angular velocity to the orbital
angular velocity at periastron ($\Omega/\omega$) is 5.83.

\begin{figure}[t!]
\plottwo{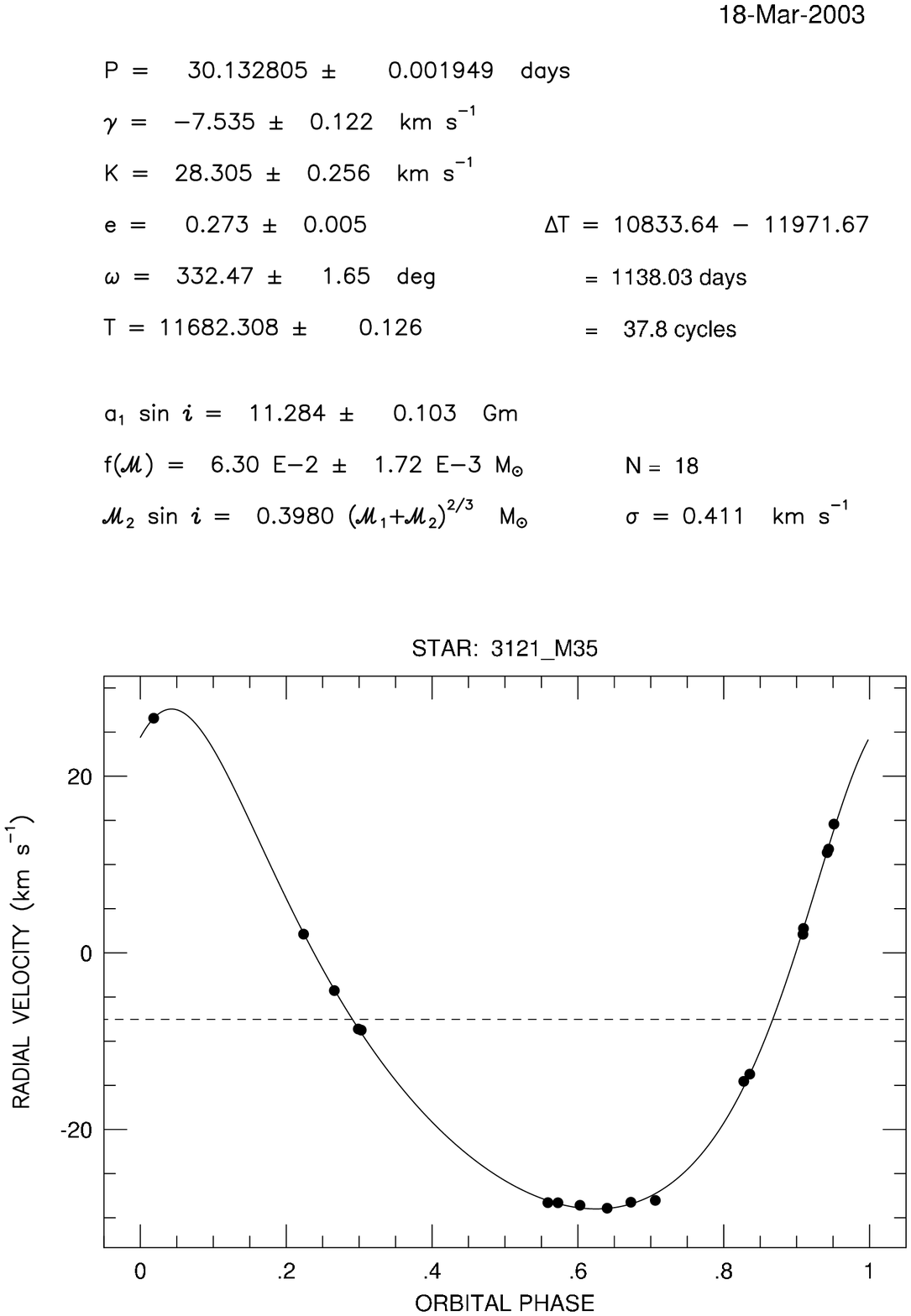}{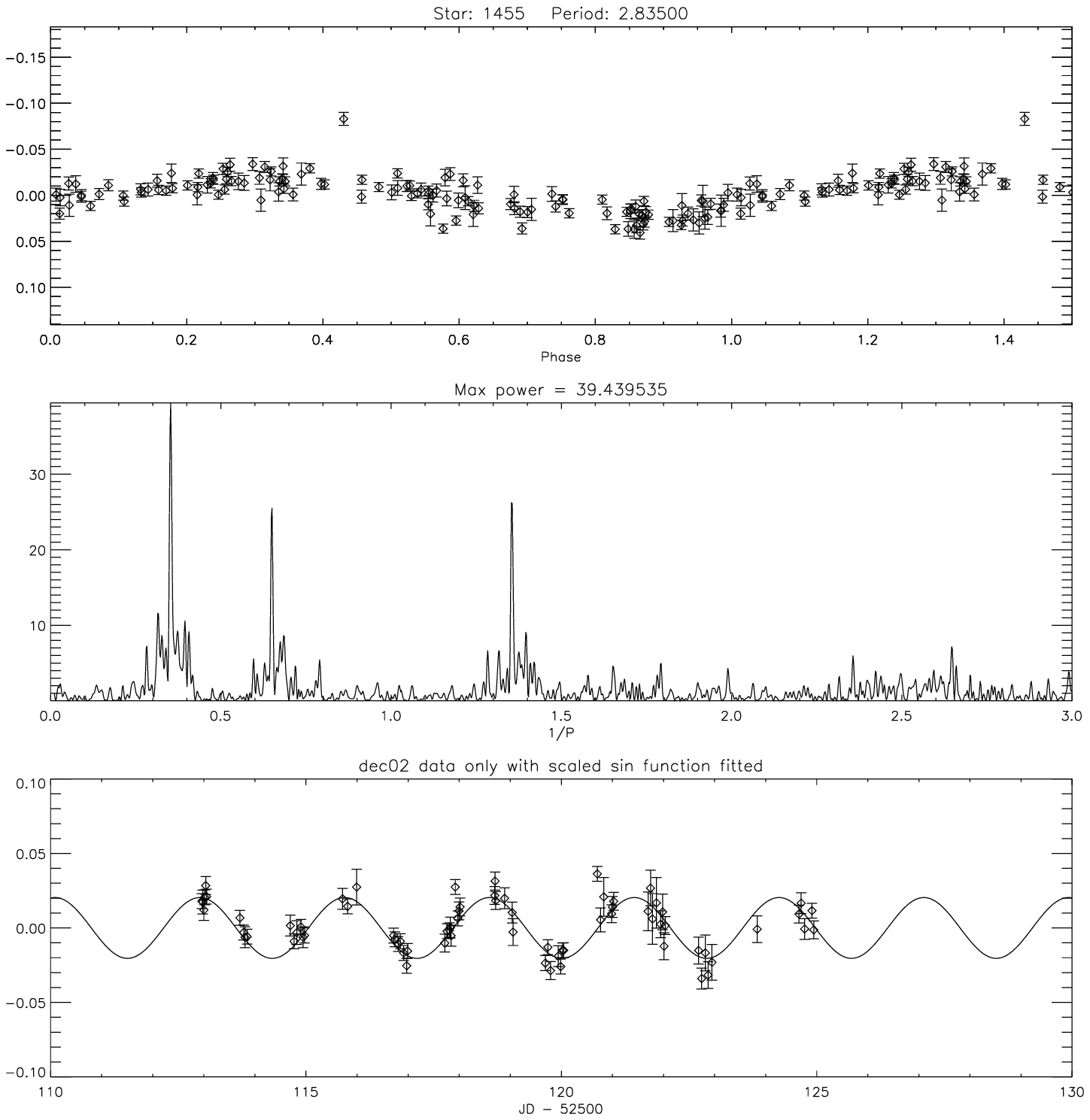}
\caption{{\bf Left:} The orbital parameters and phased radial
velocity curve for binary 3121. The best fit orbit has an eccentricity
of 0.01 and a period of 30.1 days. {\bf Right:} The photometric
light-curve for binary 3121 phased to a period of 2.8 days together
with the power spectrum in frequency space and the unphased light-curve.
A sine curve with a period of 2.8 days has been overplotted
the unphased photometric data.}
\end{figure}

Binary 701 (Figure 6): This SB1 has a highly eccentric orbit
($e = 0.54$) with a period of 8.2 days. The estimated mass
of the primary component is $0.77~M_{\odot}$. The photometric
data show a 3.6 day periodic variability in the integrated
stellar brightness of the binary. The phased photometric
light-curve is shown on the right in Figure 6. Comparing the
orbital period and the period of variability in the photometric
data might suggest that the primary star is rotating with
super-synchronous velocity. However, due to the high eccentricity
the orbital velocity at periastron is high and the derived
ratio of stellar angular velocity to the orbital angular
velocity at periastron ($\Omega/\omega$) is 0.37. The primary
in binary 701 is thus rotating at a sub-synchronous velocity. 
  
\begin{figure}
\plottwo{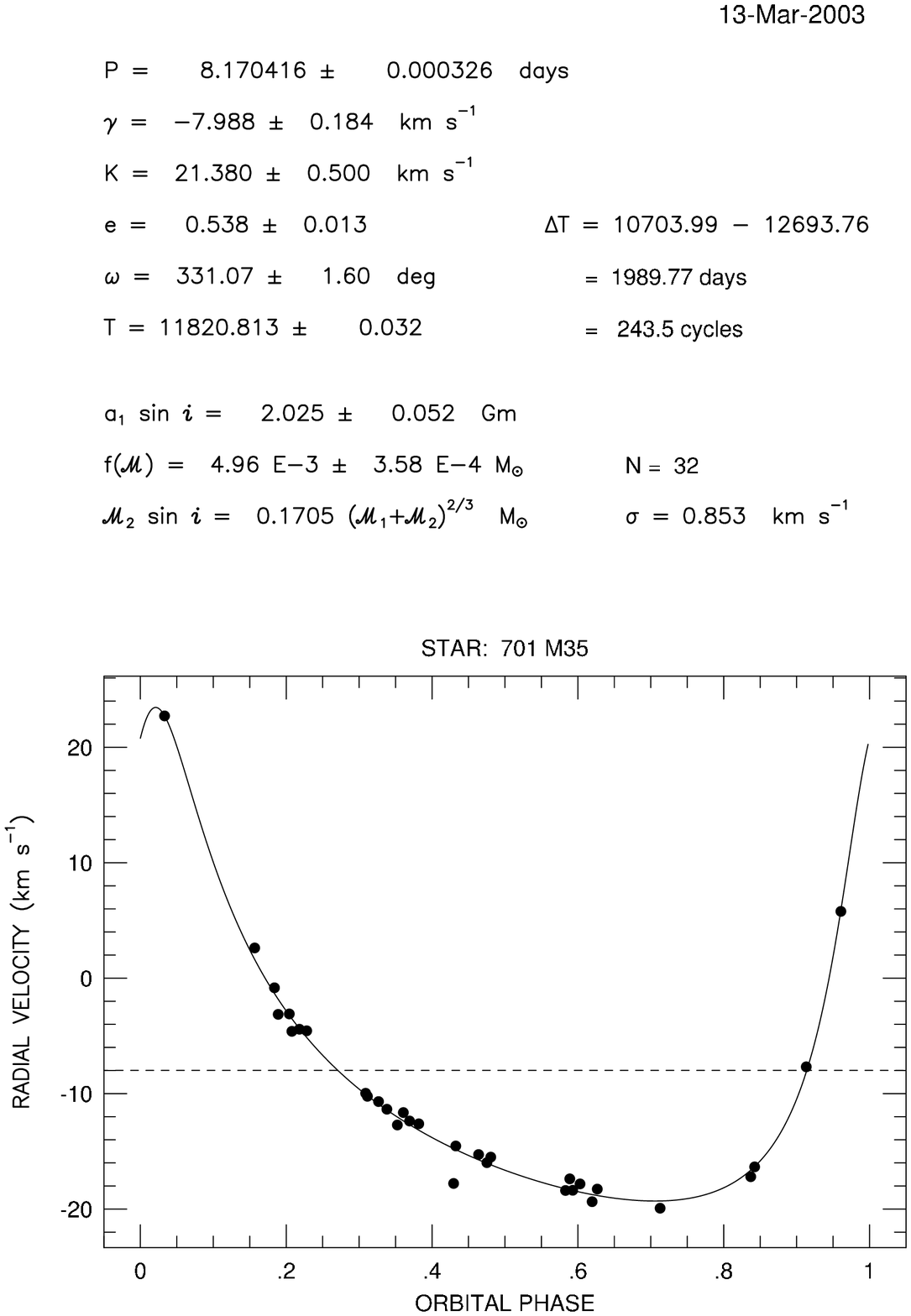}{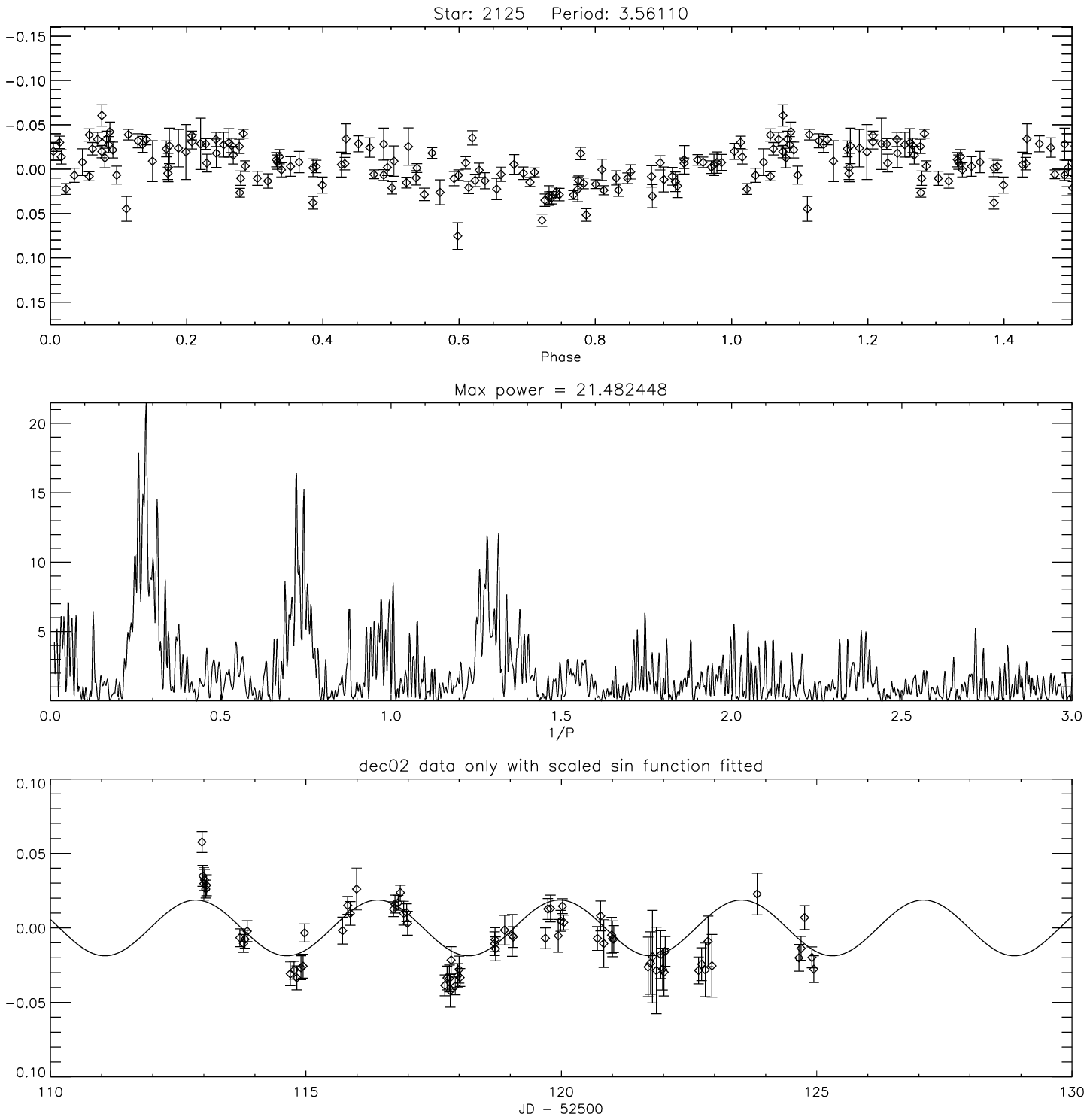}
\caption{{\bf Left:} The orbital parameters and phased radial
velocity curve for binary 701. The best fit orbit has an eccentricity
of 0.54 and a period of 8.2 days. {\bf Right:} The photometric
light-curve for binary 701 phased to a period of 3.6 days together
with the power spectrum in frequency space and the unphased light-curve.
A sine curve with a period of 3.6 days has been overplotted
the unphased photometric data.}
\end{figure}

\section{The $\Omega/\omega - \log(P)$ Diagram}

We show in Figure 7 the $\Omega/\omega - \log(P)$ diagram
with all preliminary observational results from M35 and M34.
A value of 1 for the ratio of rotational to orbital angular
velocity ($\Omega/\omega = 1$) implies synchronization or pseudo-
synchronization, whereas a value above or below 1 indicates
super- or sub-synchronous stellar rotation, respectively.
Black circles represent binary stars in M35 for which we
have secure determination of the orbital period and eccentricity,
and the rotational period of the primary star. Grey circles
represent binaries in M35 for which we are not 100\% confident
in the values for the orbital period and/or eccentricity. The 
black open square represents a binary in M34. All binary stars
shown in Figure 7 are cluster members.

\begin{figure}[ht!]
\plotone{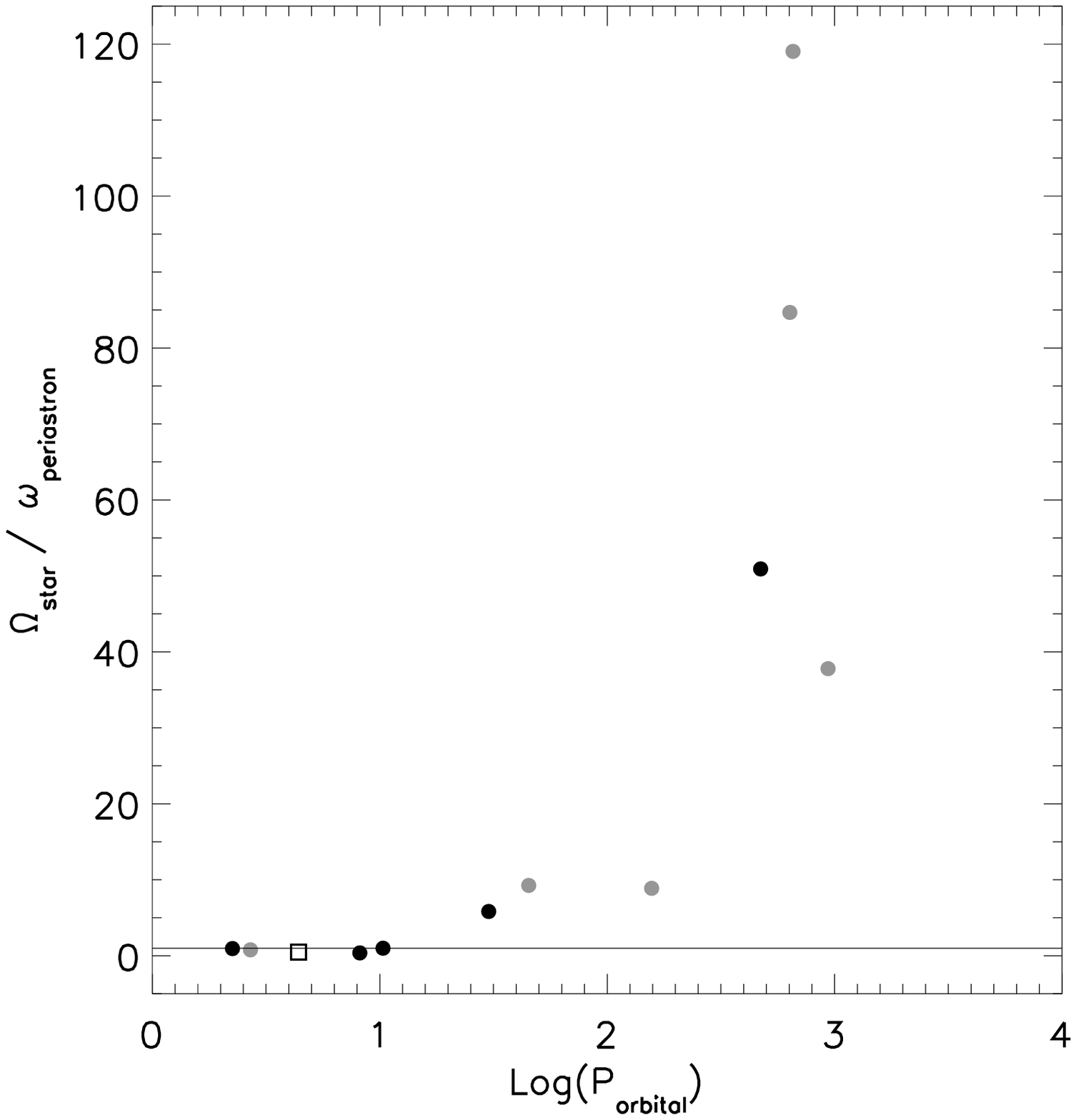}
\caption{The ratio of angular rotational velocity of the primary
star ($\Omega$) to the orbital angular velocity at periastron
($\omega$) for close detached binary stars in M35 and M34.
Black circles represent binary stars in M35 for which we have
secure determination of the orbital period and eccentricity,
and the rotational period of the primary star. Grey circles
represent binaries in M35 for which we are not 100\% confident
in the values for the orbital period and/or eccentricity.
The black open square represents a binary star in M34.}
\end{figure}

The results presented here are preliminary and the number of
binary stars with secure observational determination of orbital
and stellar angular velocities is small. Further populating the
$\Omega/\omega - \log(P)$ diagram may enable us to determine
a {\it synchronization period} at the age of the binary population.
The synchronization period will be defined as the orbital period
at which binaries at the age of the population goes from being
synchronized or pseudo-synchronized to being non-synchronized.
The synchronization period will play an important role in
constraining the efficiency and evolution of synchronization
predicted by the theories of tidal synchronization, much as
the {\it circularization period} is constraining the theories
of tidal circularization (Meibom \& Mathieu 2004; Mathieu this
volume).

Despite the small number of binary systems currently available,
some interesting individual systems offer constraints on current
theoretical models. Figure 8 show the $\Omega/\omega - \log(P)$
diagram for binaries with periods shortward of 100 days. The
coeval binaries 774 and 701 in M35 offer an intriguing challenge
to the theoretical models. Binary 774 has been synchronized and
circularized at a period of 10.3 days while binary 701 at a shorter
period of 8.2 days has not been either pseudo-synchronized or
circularized. In addition, a single binary from M34 is shown
with an orbital period of 4.4 days (black open square). The
rotational period of the primary star is measured at 8 days
(Barnes, priv. comm.). The binary orbit is circular and thus
the primary star is rotating sub-synchronously ($\Omega/\omega
= 0.48$). This is a particularly interesting result because
the timescale for tidal synchronization is thought to be
shorter than the timescale of tidal circularization
(Zahn 1977; Hut 1981).

\begin{figure}[ht!]
\plotone{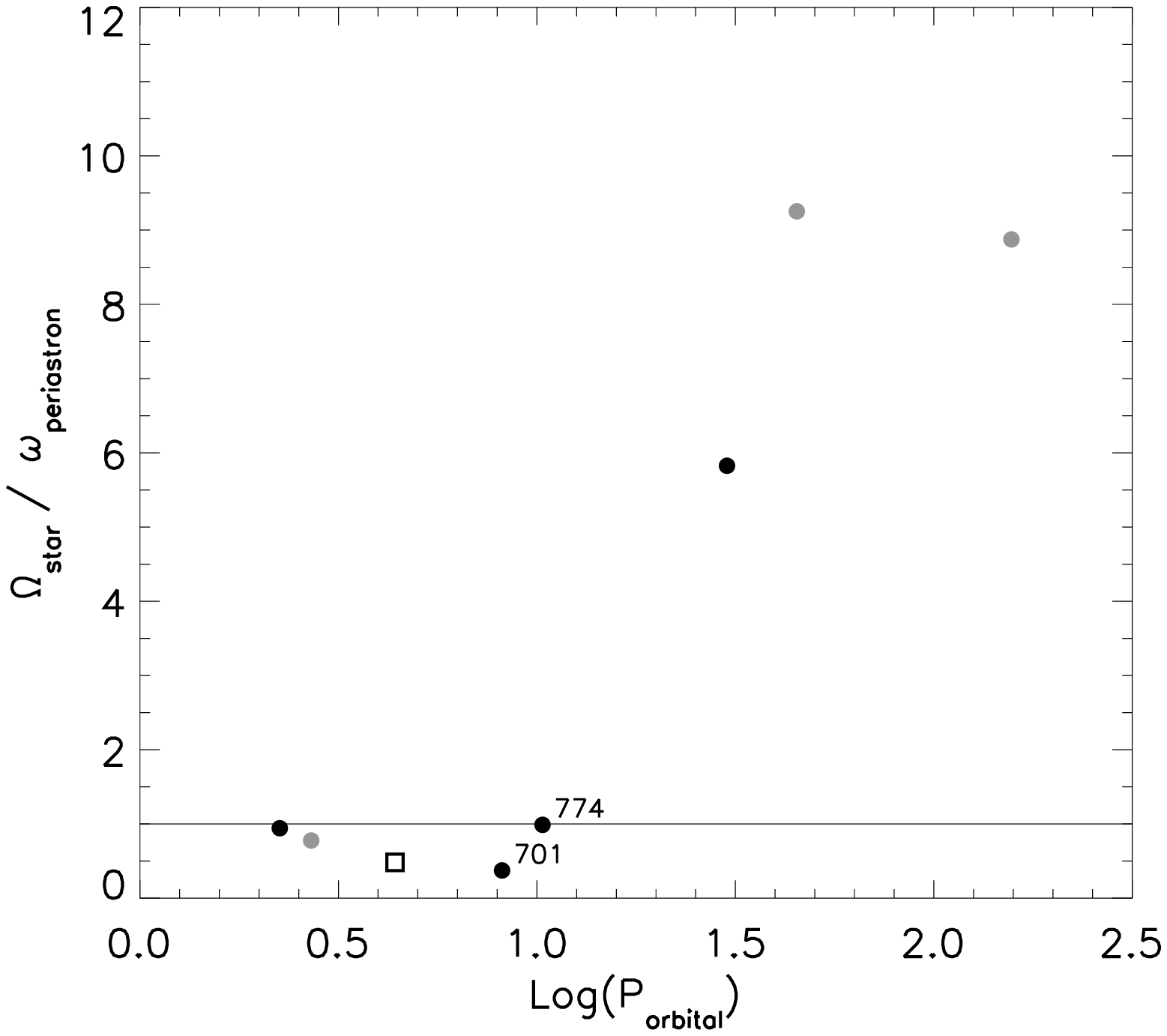}
\caption{The $\Omega/\omega - \log(P)$ diagram with focus
on binaries with orbital periods shortward of 100 days.
All symbols are the same as in Figure 7.}
\end{figure}

~\\

{\it Acknowledgments} S.M. thanks his dissertation advisor
Robert D. Mathieu. We are thankful to the University of Wisconsin
- Madison Astronomy Department, to NOAO, and to the WIYN
Telescope Director George Jacoby for the time granted on
the WIYN telescopes. We thank the WIYN Observatory staff
for exceptional and friendly support. We would like to
express our appreciation for the help from Keivan Stassun
and Sydney Barnes in planning the photometric time-series
surveys, and to Keivan Stassun for the help with the process
of deriving stellar rotation periods. We thank the organizers
for an exciting and well organized meeting in the wonderful city
of Granada. This work has been supported by NSF grant AST
97-31302 and by a Ph.D fellowship from the Danish Research
Agency (Forskningstyrelsen)
to S.M.

\end{document}